\documentclass[twocolumn,aps,showpacs,prb]{revtex4}
\usepackage{graphicx}
\usepackage{dcolumn}
\usepackage{bm}
\def \lafepo{LaFePO}

\def \hc2{$\mu_0H_{c2}$}
\def \tc{$T_{c}$}

\begin{document}

\title{Bulk superconductivity and disorder in single crystals of $\rm LaFePO$}

\author{James G. Analytis$^1$, Jiun-Haw Chu$^1$, Ann S. Erickson$^1$, Chris Kucharczyk$^1$, Alessandro Serafin$^2$, Antony Carrington$^2$,
  Catherine Cox$^3$, Susan M. Kauzlarich$^3$, H\aa kon Hope$^3$, I. R. Fisher$^1$}
  \affiliation{$^1$Geballe Laboratory for Advanced Materials and
  Department of Applied Physics, Stanford University, CA 94305, USA.}
\affiliation{$^2$H. H. Wills Physics Laboratory, University of Bristol, 1 Tyndall Ave., Bristol BS8 1TL, UK}
\affiliation{$^3$Department of Chemistry, University of California, Davis, California 95616, USA}

\date{\today}

\begin{abstract} 
We have studied the intrinsic normal and superconducting properties of the oxypnictide LaFePO. These samples exhibit bulk superconductivity and the evidence suggests that stoichiometric LaFePO is indeed superconducting, in contrast to other reports. We find that superconductivity is independent of the interplane residual resistivity $\rho_0$ and discuss the implications of this on the nature of the superconducting order parameter. Finally we find that,
unlike $T_c$, other properties in single-crystal \lafepo\, including the resistivity and magnetoresistance, can be very sensitive to disorder.\end{abstract}


\pacs{74.25.Fy, 74.25.Ha, 74.70.-b, 72.80.Ng}

\maketitle

\section{Introduction}

High-temperature superconductivity in the iron-pnictides has recently attracted considerable
attention because it is the first family of materials since the discovery of the cuprates to
exhibit critical temperatures above 40K. However, in the initial flurry of
activity, single-crystal measurements of the lower $T_c$ analogue \lafepo\, have been somewhat overlooked. Early measurements of polycrystalline samples indicated $T_c$ values in the range from 4 to 7 K,\cite{hosonoprb,hosono2,hosono,tegel,liang} with F-substitution (i.e. electron doping) \cite{hosono2} or Ca-substitution (hole doping) \cite{hosono} modestly
increasing $T_c$ values. However, more recent measurements have queried whether the
stoichiometric compound is actually superconducting, suggesting rather that oxygen deficiency
plays a significant role.\cite{cava} Initial single-crystal work revealed superconducting
transitions in the resistivity and susceptibility, but not in the heat capacity, leading the
authors to speculate that the superconductivity might be associated with an oxygen deficient
surface layer, while the bulk remains metallic to the lowest temperatures.\cite{maple} Given the
close structural and electronic similarity to the parent compounds of the higher $T_c$
oxy-arsenides, it is particularly important that a clear picture of superconductivity in this
lower-$T_c$ analog is elucidated.

\begin{figure}[tbh]
\includegraphics[width=8.0cm]{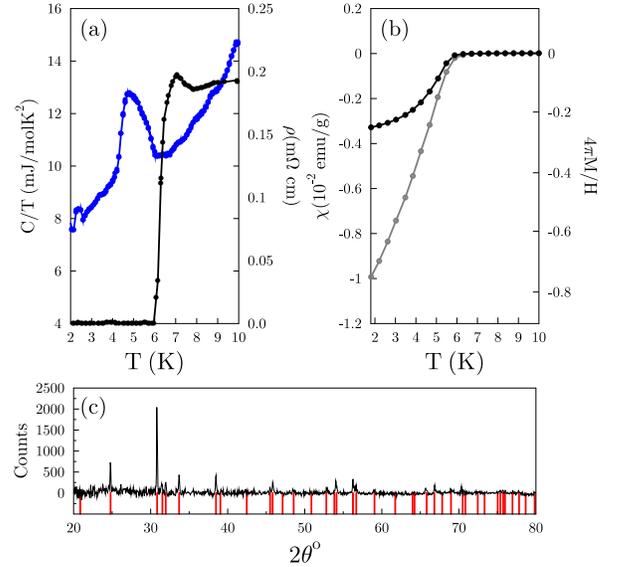}
\caption{(Color online)(a) Specific heat (left axis) of a mosaic of single-crystals and
resistivity (right axis) of a single-crystal. The heat capacity anomaly corresponds to the
disappearance of resistance. (b) The susceptibility of powdered \lafepo\, crystals measured in
an applied field of 50 Oe following zero field cooling (ZFC) and field cooling (FC) cycles. (c)
Powder x-ray diffraction pattern (black) of crystals taken from the same batch as in (b). All
peaks can be indexed to the calculated pattern (red).} \label{heatcap}
\end{figure}

\lafepo\, is interesting for reasons other than this controversy. Because of its lower $T_c$ and $H_{c2}$, quantum oscillations can be measured in relatively moderate fields.\cite{amalia} In
addition, despite the appearance of a nearly-nested Fermi surface, \lafepo\, does not appear to
exhibit any magnetism and so the properties of the {\it metallic} ground state of the pnictides
can be examined in detail. In contrast LaFeAsO has a spin-density-wave ground state at low
temperature and only exhibits superconductivity when doped, which naturally introduces disorder,
inhibiting such detailed probes of the Fermiology.

In this paper we present results of heat capacity, magnetization, resistivity and
magnetoresistance measurements of single-crystal samples of LaFePO. The evidence suggests that
these samples are stoichiometric, and we report three major results. Firstly, we find
that superconductivity is present in the bulk of the material, contrary to the findings of some
other authors.\cite{cava,maple} Second, we find the superconducting gap is uncorrelated with the interplane residual resistivity $\rho_0$. If this quantity is representative of disorder, the lack of $T_c$ suppression is consistent with $s$-wave superconductivity.\cite{anderson} Finally, in contrast
to the superconductivity itself, properties near $T_c$ can vary greatly with disorder. We
illustrate this with measurements of the magnetoresistance.

\section{EXPERIMENT}
Single-crystals of LaFePO were grown from a molten Sn flux, using conditions modified from the
original work by Zimmer and coworkers.\cite{zimmer}  Elemental and oxide precursors were placed
in alumina crucibles and sealed in quartz tubes under a small partial pressure of argon. The
growths were ramped up in temperature to 1190$^\circ C$ followed by a slow cool. We used either
La$_2$O$_3$ or Fe$_2$O$_3$ to introduce oxygen into the melt. When La$_2$O$_3$ was used, the
highest yield was obtained for a melt containing La:La$_2$O$_3$:Fe:P:Sn in the molar ratio
1:1:3:3:57, while with Fe$_2$O$_3$ the best yield was obtained from a combination
La:Fe$_2$O$_3$:P:Sn in the molar ratio 3:1:2:24. The former method yielded our largest crystals
(up to 0.7mm on a side) and with the highest RRR (residual resistivity ratio,
$\rho(300K)/\rho(0))$, but the yield was very sensitive to the packing in the crucible, the
temperature profile of the initial warm up, and the proportion of the other elements. The
temperature was raised from room temperature to 1190$^\circ C$ in 6 hours, held for a further 6
hours and then lowered to 900$^\circ C$ at a rate of 4.8 $^\circ C$/hour. Using
Fe$_2$O$_3$ yielded crystals of the right phase more consistently, although smaller (typically
0.1-0.4mm) and with a lower RRR. In this case, optimal results were obtained by heating over the
same amount of time, but the mixture was held at 1190$^\circ C$ for 18hrs, before cooling at 10
C/hour to 650$^\circ C$. The remaining liquid was decanted at the base temperature using a
centrifuge, and the crystals removed from the crucible. Excess Sn on the surface of the crystals
was removed by etching in dilute HCl followed by an immediate rinse in methanol. Both methods
resulted in multiphase growths, the principle second phase being LaFe$_2$P$_2$, which has the
ThCr$_2$Si$_2$ structure. Care must be taken to distinguish these two phases.

Crystals grown using La$_2$O$_3$ precursor as described above had RRR values up to 85, $T_c$
$\sim5.9$K and an interplane residual resistivity of $\rho_0$ = 0.04m$\Omega$cm. Crystals grown
by this technique have been used for recent ARPES \cite{donghui} measurements while crystals grown using Fe$_2$O$_3$ were used to measure de Haas-van Alphen (dHvA)
magnetization oscillations.\cite{amalia} The second method, using Fe$_2$O$_3$, with
a melt composition much closer to that used in Reference ~\onlinecite{maple}, yielded crystals
with a RRR$\sim 25$, $T_c = 6.7K$ and an interplane residual resistivity of $\rho_0 =
$0.18m$\Omega$cm.

While optimizing the synthesis conditions, batches of crystals
extracted from individual growths were checked by powder x-ray
diffraction using an SSI XPERT diffractometer.  Individual crystals
were screened by measurement of the c-axis lattice parameter. A
well-formed single crystal grown with La$_2$O$_3$ precursor and the
melt composition and temperature profile that produced the highest
quality crystals (as measured by the residual resistivity) was
selected for X-ray study. A tin deposit on the crystal was removed by
rinsing in a drop of 12 N HCl. A multiply-redundant MoK$_\alpha$ data
set of 1643 reflections ($2\theta_{max} = 63.9^\circ$) was measured on
a Bruker Apex II diffractometer. The crystal temperature was 90$\pm 1$
K. Data were corrected for absorption vith the face-indexed procedure
of SADABS\onlinecite{sheldrick} . The structure refined with
SHELX\cite{sheldrick2} to a standard R index of 0.0084, and a weighted
R index of 0.02 for all 156 unique reflections. The oxygen occupancy
is 100$\%$ within the statistical uncertainty of 1.8$\%$. The crystals
are tetragonal, space group P{\it nmm}, with cell dimensions $a =
3.941\pm 0.002 Å, c = 8.507\pm 0.005 Å$.

\begin{figure}[ht]
\includegraphics[width=8.0cm]{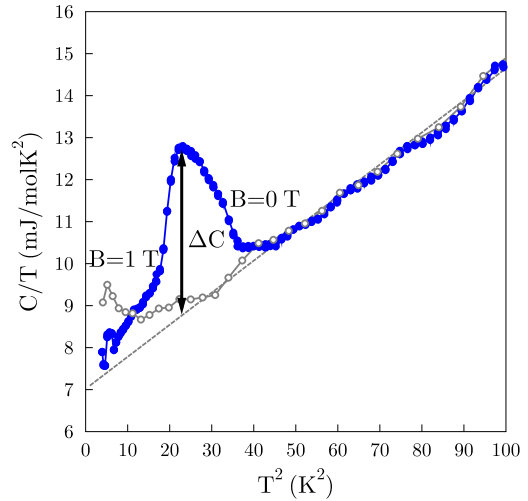}
\caption{(Color online) The specific heat C$_p$/T as a function of T$^2$ for a mosaic of
\lafepo\, crystals taken in zero field (solid circles) and in an applied field of 1 T (open
circles) with field oriented parallel to the $c$-axis. The dotted line indicates the linear fit
as described in main text which extrapolates to a value of $\gamma\sim7.0\pm0.2$ mJ$/{\rm
mol}K^2$.} \label{hcfit}
\end{figure}

Inter-plane ($c$-axis, $\rho_c$) and in-plane ($ab$-plane, $\rho_{ab}$) resistivity measurements
were made with a standard four-probe configuration and metallic contact was made using Epotek
H20E paste. The paste was annealed in air at 200$^\circ C$ for 20 minutes. Inter-plane
measurements were made on crystals as small as 0.15x0.1x0.02mm$^3$ and in-plane measurements
were made on larger crystals. However, the larger crystals tended to suffer from inhomogeneity
(evidenced by a broad superconducting transition), and/or flux inclusions, and the c-axis data
were generally better. Resistivity and heat capacity measurements were performed on a Quantum
Design PPMS (14T). Due to the small size of the crystals, the heat capacity was measured for a
mosaic of around 100 crystals grown from a melt containing La$_2$O$_3$, with a total mass of
800$\pm 50\mu g$, each oriented such that the $ab$-plane was parallel to the platform. Crystals
from a separate batch were powdered, half of which was checked for phase purity by powder x-ray
diffraction ( Figure \ref{heatcap} (c)). Susceptibility measurements were made for the other
half of this powder sample using a Quantum Design MPMS 5 Superconducting Quantum Interference
Device magnetometer in a range of applied fields.
\begin{figure}
\includegraphics[width=8.0cm]{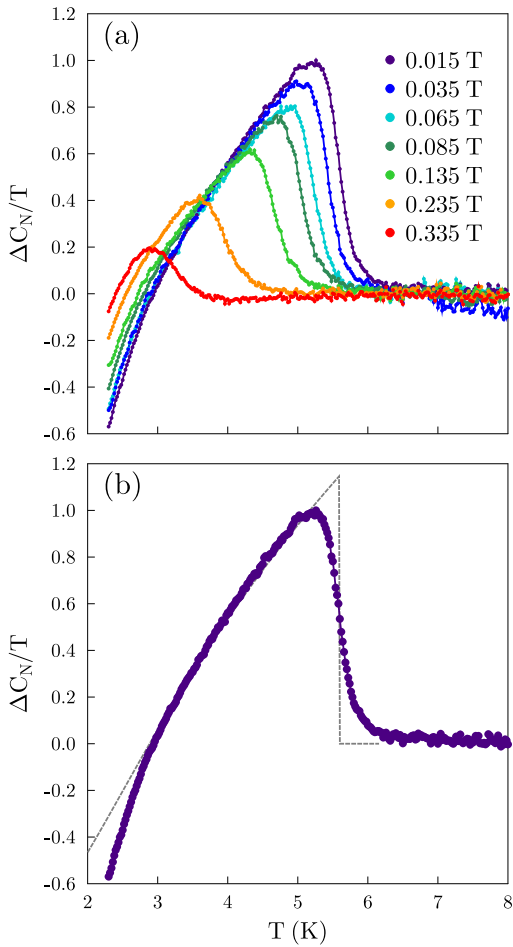}
\caption{ Heat capacity data on a single-crystal ($\sim$5$\mu g$) measured using the AC technique giving the relative change of the the heat capacity $C$ from the normal state $C_N$ (in arbitrary units), where $\Delta C=C-C_N$. In (a) we show the field dependence of the heat capacity anomaly, approximately agreeing with the suppression of \tc\, extracted from resistivity (see Figure \ref{anomvar}). In (b) we show data fitted with weak-coupling BCS theory assuming an isotropic $s$-wave order parameter. The data agrees very well to the theory.}
  \label{tonyfig}
\end{figure}

\begin{figure}
\includegraphics[width=8.0cm]{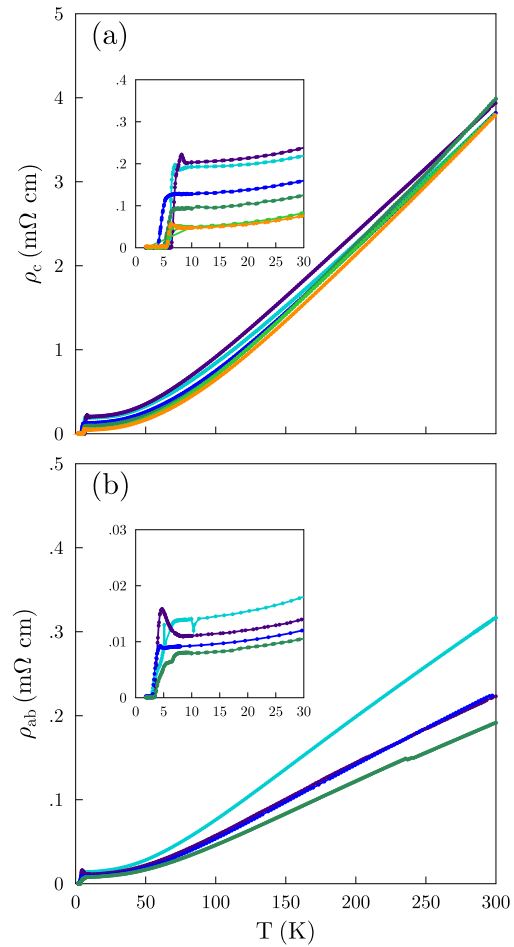}
\caption{(Color online) (a) Interlayer ($c$-axis) resistivity for \lafepo\, crystals
  grown using Fe$_2$O$_3$ (blue and green curves) and La$_2$O$_3$ (all other curves). The room
  temperature
  values are in approximate agreement for either method but the
  residual resistance $\rho_0$ shows significant variability (inset). (b)
  In-plane ($ab$-plane) resistivity. The room temperature
  value shows more variability than the interplane resistivity,
  primarily due to difficulties in finding homogeneous samples large
  enough to mount for in-plane measurements. This was often manifested
  in a broader superconducting transition (inset).}
  \label{resop}
\end{figure}

Measurements of heat capacity were also made on a single crystal using
an AC method.\cite{carrington} The sample
with dimensions 140$\times$220$\times$30$\mu$m$^3$ (with mass $\sim
5\mu$g) we attached with GE varnish to a 12$\mu$m diameter
Chromel-Constantan thermocouple, and heated with light at a frequency
of 16Hz. The size of the temperature oscillations is inversely
proportional to the heat capacity of the sample plus the addenda
(varnish plus thermocouple).  Although with this method it is possible
to measure a very small single-crystal it is not possible to determine
the absolute value of the specific heat.  The heat capacity was
measured as a function of field from -0.4T to 1T with $B\parallel c$, in a 14T
superconducting magnet. To account for the residual field in our
magnet the actual field was determined from the field which produced
the maximum field in a $T_c$ versus $H$ plot (the offset was 0.065 T).

\section{RESULTS}

Heat capacity data for the mosaic of single-crystal samples described above are shown in Figure
\ref{heatcap}(a) together with representative resistivity data for a single-crystal from a
separate batch. The total heat capacity of this mosaic is extremely small due to the small mass,
making measurements of the relaxation time particularly challenging. Despite these difficulties, a clear
superconducting anomaly is evident at 5.9 K, corresponding to the temperature at which the
resistivity disappears. The superconducting state can be suppressed by an applied field greater
than $\mu_0H_{c2}$, leaving only the normal state contribution to the heat capacity (Figure
\ref{hcfit}). Data between approximately 5 and 10 K can be fitted using $C_v/T=\beta T^2+
\gamma$ where $\beta$ and $\gamma$ are the lattice and electronic heat capacity coefficients
respectively. This yields an estimate of $\gamma = 7.0\pm 0.2\,{\rm mJ/mol}K^2$ where one mole
refers to the formula unit LaFePO. This value is similar to that obtained from values extracted
by other authors for polycrystalline \cite{cava,hosono} and single-crystal\cite{maple} samples,
though care should be taken on the definition of a mole (the value is doubled if one mole
refers to the unit cell). Data on the low-temperature side of the superconducting transition are
less reliable due to the small value of the heat capacity and the experimental difficulties
measuring the relaxation time as described above. Due to these difficulties, the data does not appear to conserve entropy. Combined with the
rounding of the superconducting anomaly seen in the data, it is difficult to obtain an accurate
estimate of the heat capacity jump,  but we can obtain a conservative lower bound by comparing
the peak of the transition with the normal state extrapolation (vertical arrow in Figure \ref{heatcap}). This
results in an estimate of the normalized heat capacity jump $\Delta C/\gamma T_c \approx
0.6\pm0.2$, which is greater than Kamihara {\it et al.}'s estimate based on measurements of
polycrystalline samples,\cite{hosono2} and is less than (about 40$\%$) the BCS result $(\Delta
C/\gamma T_c)_{\rm BCS} = 1.43$. The actual value is likely somewhat larger than this lower
bound.
Susceptibility measurements for a sample consisting of several powdered single-crystals are
shown in Figure \ref{heatcap} (b) for an applied field of 50 Oe following both zero field cooling
(ZFC) and field cooling (FC) cycles. Perfect diamagnetism ($4\pi M/H = -1$, shown on the right
axis in Figure 1b) corresponds to a value of -1.32$\times 10^{-2} emu/g$. Assuming that the
particle size is larger than the penetration depth ($\sim 0.2\mu m$\cite{malone}), the FC value
($\sim-0.32 emu/g$) provides the better estimate of the superconducting volume fraction, giving
a lower bound of approximately 24$\%$. This likely substantially underestimates the actual value due to flux pinning effects.

\begin{figure}
\includegraphics[width=8.0cm]{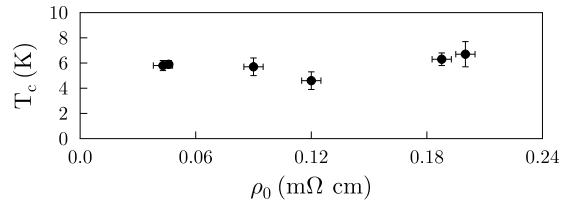}
\caption{Plot showing the dependence of the critical temperature $T_c$ on the residual resistivity $\rho_0$ extracted from the low temperature normal state of \lafepo. The higher residual resistivities correspond to a ratio of the coherence length to mean free path $\xi_{ab}/l\sim 0.17-0.25$, and so a partial, systematic suppression in $T_c$ is expected for an anisotropic order parameter.}
  \label{agplot}
\end{figure}

Heat capacity data for a single crystal, obtained using the AC technique, are shown in Figure \ref{tonyfig}. For a field of $> 0.47$T the superconducting anomaly was suppressed to
below T=2.3 K so we used this data as an estimate of the normal state
heat capacity $C_N$ (including addenda), and subtracted this from the
data in zero field.  The superconducting anomaly is
now clearly visible, with $T_c$ (mid point of the increase in $C$ of
T=5.6 ~K and width (10-90$\%$) of 0.4K.  In Figure \ref{tonyfig} (a), we show the field dependence of the anomaly in $C$.  The
decrease in \tc\, corresponds to $dT_c/dH=7.2\pm0.1$K$/$T. Also shown Figure \ref{tonyfig} (b) is the
temperature dependence calculated from the weak-coupling BCS theory
assuming an isotropic $s$-wave order parameter.  Apart from the data
very close to \tc\, or for T$<$2.8K the theory fits the data very
well. The departure below $\sim$2.8K probably arises from the field
dependence of the thermocouple thermopower and field dependence of the
addenda which we have not attempted to correct for.

In the strong coupling Eliashberg theory the slope near \tc\, normalised to
the jump at $T_c$ $R=\frac{T_c}{\Delta C}\frac{\partial C}{\partial
T}|_{T_c}$ is indicative of the coupling strength.\cite{atkis} The fact that our data fit well the weak
coupling BCS form shows that LaFePO is in the weak coupling limit (note
however this does not prove the order parameter is necessarily
isotropic).

\begin{figure}
\includegraphics[width=8.0cm]{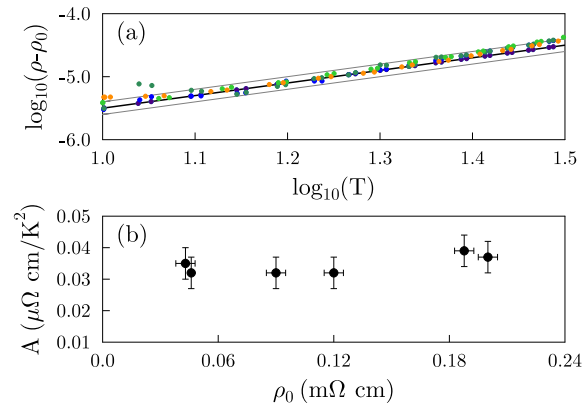}
\caption{(Color online) (a) c-axis resistivity data from Figure 3(a) replotted on a logarithmic
scale. Plotted this way, all data fall on the same curve. The black line shows the
  value $A\sim3.2\times10^{-5}$m$\Omega$cm$/K^2$, while the grey lines
  illustrate our error bars in $A$. (b) The coefficient $A$ extracted
  from fits of the resistivity data to $\rho=\rho_0+AT^2$, which is
  independent of disorder.}
  \label{tsq}
\end{figure}

In Figure \ref{resop}(a) we show the inter-plane (c-axis) resistivity $\rho_c$ for five crystals
grown from slightly different melt conditions. Absolute values of the room temperature
resistivity are in excellent agreement, and the principal difference between the samples is the
residual resistivity $\rho_0$. A small upturn is seen in the resistivity of some but
not all samples, just above $T_c$. $T_c$ values are also in excellent agreement, and show no
correlation with the residual resistivity (see Figure \ref{agplot}). Similar data for in-plane resistivity measurements $\rho_{ab}$ are shown in
Figure \ref{resop}(b). In contrast to the $c$-axis data, in-plane measurements tend to exhibit
both broader transitions and also a wider variation in the absolute value of the resistivity. As
described above, this difference can be ascribed to the use of larger crystals for the in-plane
measurements, which are consequently more susceptible to defects, flux inclusions and, due to
the malleability of the crystals, macroscopic distortion. The anisotropy of the interplane to
in-plane resistivity $\rho_{c}/\rho_{ab}$ is in the range from 13 to 17, in approximate
agreement with band structure calculations.\cite{kamihara,singh,lebegue}

Neglecting the low-temperature upturn seen in some samples, resistivity data below 25K can be
well-fit by the standard Fermi liquid behavior, $\rho=\rho_0+AT^2$, where $\rho_0$ is the
residual resistivity.\cite{upturnnote} To confirm the reliability of this fitting,
log($\rho-\rho_0$) vs log($T$) is shown in Figure \ref{tsq}(a) for the c-axis data, which
reveals that each data set not only has a gradient of $2\pm0.3$, but also falls on the same
value of $A\sim3.2\pm0.5\times 10^{-5}$m$\Omega$cm$/K^2$ (Figure \ref{tsq}(b)). We note in
passing, that when $A$ is combined with our value of $\gamma$ we find a Kadowaki-Woods ratio of
$A/\gamma^2=65\pm15\times 10^{-5} \mu\Omega $cm(mol K/mJ)$^2$.\cite{kadowaki}

\begin{figure}
\includegraphics[width=8.0cm]{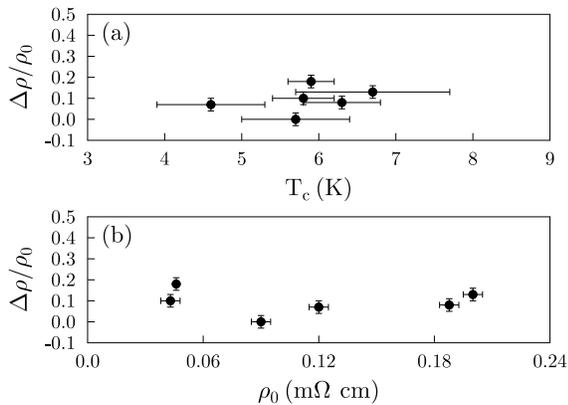}
\caption{(a) The magnitude of the interplane resistivity upturn
  $\Delta\rho_c=\rho_p-\rho_0$ normalized to the residual resistivity
  $\rho_0$, where $\rho_p$ is the peak resistivity of the upturn. There does not appear to be a correlation with $T_c$. (b)
  The magnitude of the interplane resistivity upturn $\Delta\rho_c$ as
  a function of the interplane residual resistivity $\rho_0$, which
  also shows no correlation.}
  \label{delr}
\end{figure}

As mentioned above, a small resistivity upturn is observed in some, but not all, samples just
above $T_c$. This effect is seen for both current orientations (i.e. for $\rho_{ab}$ and
$\rho_c$). Significantly, the size of the upturn is not correlated with either $T_c$ (Figure
\ref{delr}(a)), or with the absolute value of the residual resistivity $\rho_0$ extrapolated
from higher temperatures (Figure \ref{delr}(b)).  We return to the field dependence and origin
of this feature below.

\begin{figure}
\includegraphics[width=8.0cm]{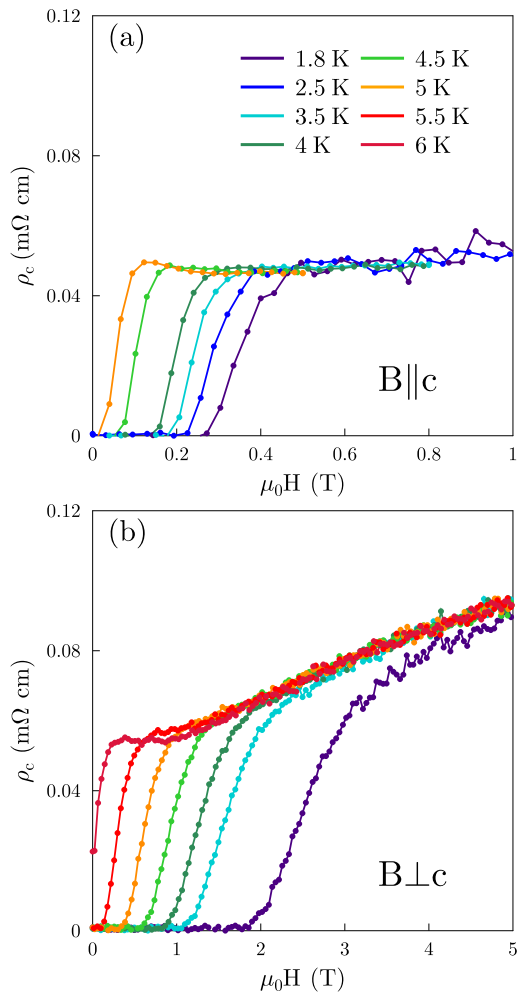}
\caption{ (Color online) Interlayer (c-axis) resistivity as a function of magnetic field
  for (a) field parallel and (b) orthogonal to the $c$-axis for temperatures from 1.8 to 6.0 K
  (see legend in upper panel). The
  anisotropy in $H_{c2}$ is approximately 6 in this case, though other
  samples show an anisotropy of up to 15.} \label{opmagres}
\end{figure}

\begin{figure}
\includegraphics[width=8.0cm]{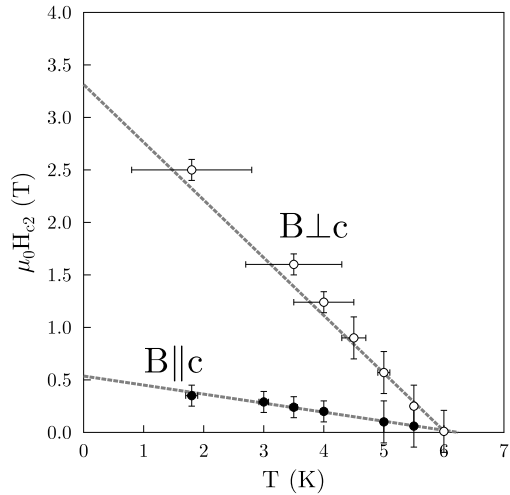}
\caption{The critical field \hc2 for fields oriented parallel (solid symbols) and perpendicular
(open symbols) to the c-axis. Data points were determined from resistivity data as described in
the main text. Error bars indicate 10 and 90 $\%$ of the superconducting transition.}
\label{hccomp}
\end{figure}

Representative data showing the variation in the c-axis resistivity as a function of applied
field at various temperatures are shown in Figure \ref{opmagres} for fields applied (a) parallel
and (b) perpendicular to the c-axis. From data like these we are able to extract $\mu_0H_{c2}$,
choosing for simplicity to define $\mu_0H_{c2}$ in terms of the midpoint of the resistive transition, and using 10 and 90$\%$ of the transition to determine the error bars.  Plotting \hc2\ extracted in this manner, as shown in Figure \ref{hccomp} and extrapolating to $T=0 K$, we determine $\mu_0H_{c2}^{\perp}(0) \sim 0.6 T$ and $\mu_0H_{c2}^{\parallel}(0) \sim 3.5 T$, giving an anisotropy of $\sim6$. There is some variation in this value, perhaps due to the sensitivity of $H_{c2}^{\parallel}$ to orientation, but most samples yield an anisotropy between 6 and 15, similar to that observed by other authors.\cite{maple}

\begin{figure}
\includegraphics[width=8.0cm]{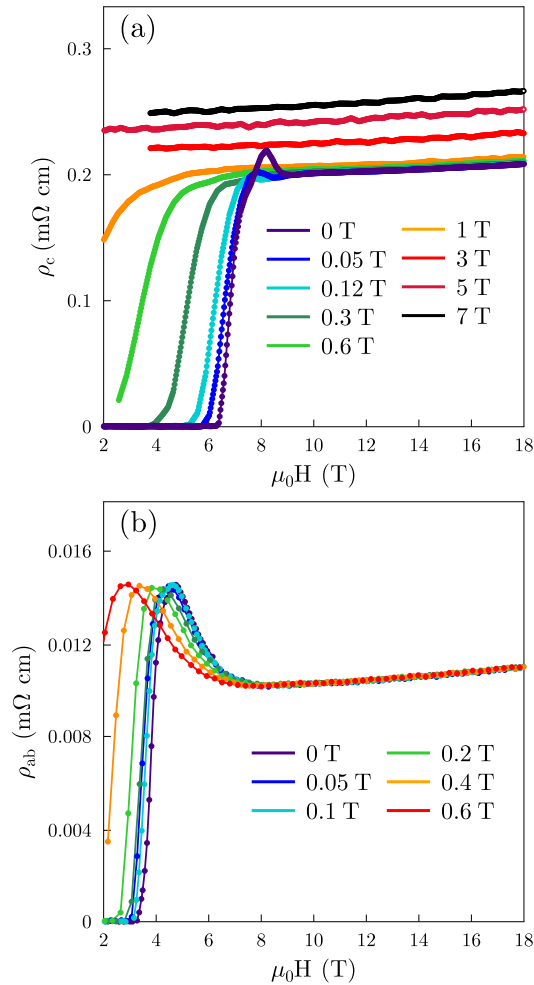}
\caption{ (Color online) Temperature sweeps at constant field for field directed
  parallel to the $c$-axis illustrating the effect of applied fields on the resistivity upturn
  observed at low temperatures for some samples. In (a) we show the interplane resistivity
  for a sample with a small upturn and in (b) we show the in-plane
  resistivity for a sample with a pronounced up-turn. In both cases
  the anomaly is rapidly suppressed by magnetic field, though clearly
  the more pronounced the upturn, the higher the field required for
  the suppression.} \label{anomvar}
\end{figure}

The resistivity upturn which is observed in several samples is suppressed by an applied field,
resulting in a pronounced ``hump" feature in the magnetoresistance $\rho(H)$ for fields just
above $H_{c2}$\, (see Figure \ref{opmagres} ). The magnitude of the resistivity upturn in zero
field, and hence the associated negative magnetoresistance, varies between samples grown under
different conditions, and to a lesser extent even between samples from the same batch. This
variation is illustrated in Figure \ref{anomvar} which show representative c-axis and in-plane
magnetoresistivity data for samples with small (panel (a)) and large (panel (b)) upturns
respectively. The field is oriented parallel to the c-axis in both cases to allow direct
comparison. For the sample shown in panel (a), the anomaly is suppressed by fields as small as
$0.3 T$, whereas for the sample shown in panel (b) it survives to fields of more than twice this
value. In all cases, once the anomaly is completely suppressed, a positive magnetoresistance
appears which is approximately linear with field in the high-field limit.

\begin{figure}
\includegraphics[width=8.0cm]{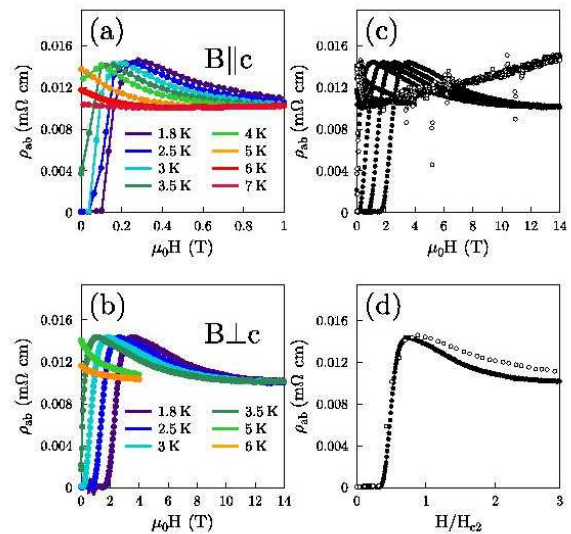}
\caption{ (Color online) In-plane magnetoresistance for a sample with a pronounced
  upturn. At low fields the magnetoresistance is negative until the
  anomaly is completely suppressed. (a) illustrates the suppression
  when field is applied in-plane while (b) shows that the anomaly
  extends over a much wider field range when the field is applied
  orthogonal to the $c$-axis. (c) Illustrates both these data set on
  the same plot for direct comparison and (d) shows the data at 1.8 K scaled by $\mu_0H_{c2}^{ab}$
  for the in-plane data and  $\mu_0H_{c2}^{c}$ for the inter-plane data.}
  \label{fsanom}
\end{figure}

Significantly, the suppression of the resistivity upturn is strongly dependent on the
orientation of the applied field with respect to the $c$-axis. For example, for the sample shown
in Figure \ref{fsanom} (a-c), the resistivity upturn is suppressed for fields exceeding a modest 1 T when
applied parallel to the $c$-axis, but is only suppressed for fields exceeding 12 T when directed
perpendicular to the $c$-axis. Although the absolute values of the fields at which the
resistivity upturn are suppressed vary between crystals depending on the magnitude of the
zero-field upturn, this anisotropy is always observed. This anisotropy appears to mirror the
anisotropy in $\mu_0H_{c2}$, such that if magnetoresistance data for the two orientations are plotted
as a function of field scaled by the appropriate value for $\mu_0H_{c2}$ for that orientation, the
data lie almost on top of each other (Figure \ref{fsanom} (d)). It is not clear whether this is
coincidental, or reflects a deeper inter-relation between superconductivity and the origin of
the resistivity upturn.

\section{DISCUSSION}
The results summarized in Figure \ref{heatcap}-\ref{tonyfig} provide the first evidence of superconductivity
in single-crystals of  stoichiometric \lafepo, manifested in a bulk thermodynamic
transition in both the magnetization {\it and} the heat capacity.  In contrast, despite the
observation of bulk superconductivity in polycrystalline samples by several other groups,\cite{hosono,liang,tegel} McQueen et al. have recently suggested that the stoichiometric
compound is actually non-superconducting, and that the onset of superconductivity is perhaps associated
with the presence of oxygen deficiencies, among other possibilities.\cite{cava} However, the absence of superconductivity in \lafepo\, would be surprising given that there is nothing anomalous in the density of states\cite{lebegue} of the stoichiometric compound, no evidence of a competing phase transition, and doping with either holes or electrons results in superconductivity.\cite{hosonoprb} Using crystals grown by a similar
technique to our own, Hamlin et al. observed a substantial diamagnetic susceptibility, but no
anomaly in the heat capacity, causing them to suggest the possibility that their crystals have a thin surface layer of oxygen-deficient superconducting material surrounding a stoichiometric
non-superconducting center.\cite{maple} Refinements of single-crystal x-ray diffraction
data for our crystals indicate full site occupancy, with a standard uncertainty in the oxygen
content of only 1.8$\%$. This is not the best probe of the oxygen content, but several other
pieces of evidence suggest that oxygen deficiency, or at least variation in oxygen deficiency,
does not play a significant role in these crystals. Specifically, neither the $T^2$ coefficient
of the resistivity nor $T_c$ itself (Figures \ref{tsq} and \ref{agplot} respectively), show any significant variation with the residual resistivity, in contrast to what would be expected if variation in the oxygen content were
responsible for the different residual resistivity values if this were the principle
source of disorder in these samples. Furthermore, these values do not depend on details of the
synthesis conditions, including whether oxygen is introduced to the melt using Fe$_2$O$_3$ or
La$_2$O$_3$, and even the relative concentration of these precursors in the melt. These
observations demonstrate that there is remarkably little variation in oxygen deficiency between
crystals grown by this technique. In addition, the extremely high residual resistivity ratios,
up to 85, indicate that the material is in fact exceptionally well-ordered. Finally,
analysis of recent dHvA data taken for samples grown by this technique indicate that the
electron and hole pockets are almost balanced, in agreement with the anticipated
result for stoichiometric material,\cite{dhva_comment} with a mismatch that would correspond to a doping smaller than the error of our x-ray diffraction refinements.

Experiments involving annealing in an oxygen atmosphere may be revealing, and though these are in progress one will always have the problem of establishing by how much the material has departed from perfect stoichiometry, which is a difficult measurement for such small crystals. Thus we cannot definitively prove the absence of oxygen deficiency from the present measurements, and in fact the breadth of the specific heat anomaly in Figure \ref{heatcap} may be indicative of slight variation in stoichiometry even within a single sample. However, given the small departure from stoichiometry that our errors allow, together with the above observations, it seems extremely unlikely that the stoichiometric compound is non-superconducting. Furthermore, the absence of other competing phases suggests that superconductivity is the natural low temperature ground state. In this case, the absence of a superconducting anomaly in the heat capacity data of Hamlin et al. likely reflects similar difficulties that we experienced in the measurement of such small sample masses. It is difficult to speculate on the reason for the absence of superconductivity in the polycrystalline samples of McQueen et al.\cite{cava} but we note that the resistivity of those samples exhibits a substantial low-temperature upturn. We return to the possible significance of this below.

From the range and variety of samples measured in this work, we are also able to consider the
role of disorder on the superconductivity of \lafepo. Even though these materials are
anisotropic, the interplane transfer integral for the electron pockets is relatively large\cite{amalia} and so the
interplane residual resistivity should be a good measure of the number of defects in a given
crystal, though this argument is weaker for the hole pockets which have little warping in the $k_z$ direction.\cite{powell} Assuming that this is the case, the plot of $T_c$ vs $\rho_0$, as shown in Figure \ref{agplot}
illustrates that \tc\, is, within our uncertainties, uncorrelated with disorder. For other high-temperature superconductors the absence of $T_c$ suppression has been attributed to the smallness of the coherence length\cite{radtke} $\xi_{ab}$. From the present measurements $\xi_{ab}$ can be estimated from the relation $\mu_0H_{c2}^{\perp}=\Phi_0/2\pi\xi_{ab}^2$, giving $\xi_{ab}\sim0.03\mu m$. The mean free path can be estimated from a quasi-two-dimensional Drude model $\sigma_{\parallel}=(e^2/h)\times k_Fl/c$, which yields a different conductivity for electron and hole pockets. Using the results of Reference \onlinecite{amalia}, we estimate the electron and hole pockets to have a mean free path $l\sim1700\AA$ and $l\sim1200\AA$ respectively for the dirtiest samples, approximately in agreement with estimates from de Haas van Alphen oscillations.\cite{amalia} For the electron (hole) pockets this suggests the ratio $\xi_{ab}/l$ has shifted from a value of $\sim 0.04$ ($0.06$) for the cleanest samples to $\sim 0.17$ ($0.25$) for the dirtiest samples. If the disorder were magnetic, a partial, systematic suppression of $T_c$ should manifest for any pairing symmetry according to
the relationship first calculated by Abrikosov and Gor'kov.\cite{abrikosov,radtke} Similarly, $T_c$ for anisotropic $d$-wave or $p$-wave superconductors in the weak coupling limit should be suppressed due to the inevitable phase mixing of the order parameter, which should be noticeable by this value of $\xi/l$.\cite{powell,mackenzie,radtke} Conversely, according to
Anderson's theorem,\cite{anderson} an $s$-wave superconductor should be robust against
non-magnetic disorder due to the ability of the quasiparticle states to form time reversed
pairs. The data in Figure \ref{agplot} suggests that the predominant form of disorder is therefore non-magnetic. More importantly, if the dominant scattering is strong enough to scatter across the Brillouin zone, then this plot also illustrates that the order parameter mediating superconductivity is likely to be $s$-wave, an observation consistent with recent measurements of penetration depth\cite{malone} on SmFeAsO$_{1-x}$F$_x$. However, if the scattering is small angle, or only occurs intra-Fermi surface, then an $s$-wave order parameter which reverses sign between Fermi surfaces centered at the $\Gamma$ and $M$ points may also be possible, as proposed by Mazin {\it et al}.\cite{mazin} On the other hand, if the disorder is inhomogeneous on scales of order $\xi$, then it is possible than an anomalously clean path will yield a resistive transition $T_c$ which is essentially disorder independent, though the superfluid density would be rapidly suppressed. This behavior may be evident in the heat capacity, and the breadth of the anomaly shown in Figure \ref{heatcap} may be indicative of just such inhomogeneity.\cite{kivelson} Naturally, the above discussion relies on the interplane resistivity being an accurate measure of the in-plane scattering rate, which may be true for the electron pockets due to their significant $c$-axis dispersion, but less likely for the hole pockets.

We now turn the discussion to the origin of the resistivity upturn observed in some crystals for
temperatures just above \tc. Similar upturns are observed in several other classes of
unconventional superconductors including many of the cuprates,\cite{cho} several
organics\cite{zuo,taniguchi} and other polycrystalline oxy-arsenides.\cite{riggs} However, in
all these classes of compounds one observes an enhancement of the anomaly with an applied
magnetic field, whereas in the present case we observe a suppression of the anomaly with field.
In addition, we observe the upturn in some but not all of our samples, implying that this
behavior is intimately linked  with the presence of disorder. Finally, the anomaly is observed
regardless of the current orientation, whereas in other correlated materials the effect is only observed when the current
is passed along the $c$-axis.\cite{cho} In highly anisotropic materials with Josephson coupled planes, one expects dramatic differences
in the behavior of each current orientation in field. Specifically, near $T_c$ in-plane currents are
dissipated by the motion of pancake vortices under the Lorentz force while inter-layer currents are
dissipated by the Josephson coupling between the layers, giving rise to a broad
anomaly in $\rho_c$.\cite{cho} We thus rule out such coupling as the source of the
anomaly.

Given the relatively large concentration of Fe in the melts used to grow these crystals, it is
not unreasonable to suggest that the resistivity upturn might be associated with Kondo
scattering from magnetic impurities. However, the strong anisotropy of the upturn's suppression
with field, together with the fact that samples exhibiting the upturn do not show a dramatically suppressed \tc\, convincingly excludes this possibility. Another possible scenario for the
origin of the resistivity anomaly is weak localization due to elastic scattering defects,
possibly also explaining the rapid suppression of the anomaly with field. Furthermore, in
quasi-two-dimensional systems, there can be a significant anisotropy in the field-induced
suppression of weak localization.\cite{mauz} However, the upturn does not scale with the
residual resistivity $\rho_0$ (as shown in Figure \ref{delr} (b)), indicating that the effect
does not correlate with the concentration of elastic scattering centers.

An intriguing alternative hypothesis for the origin of the resistivity anomaly is scattering
from spin fluctuations associated with a competing magnetic phase. Near perfect nesting of
electron and hole pockets evidenced from dHvA measurements \cite{amalia} certainly raises the
question as to whether the spin density wave (SDW) observed in the isoelectronic compound LaFeAsO might be stabilized in LaFePO. In contrast, recent NMR results for La$_{0.87}$Ca$_{0.13}$FePO indicate the presence of {\it ferromagnetic} fluctuations,\cite{Nakai} perhaps indicating that
$q \sim$ 0 nesting between concentric FS sheets dominates for such high doping levels. The same
study also revealed anomalous behavior of 1/$T_{1}T$ below $T_c$ for that material, indicating
unusual spin dynamics in the superconducting state, though the origin of this effect is not yet
clear. While the evidence points towards stoichiometric LaFePO being superconducting, one can
speculate that, for example, magnetic impurities, or even disorder-induced variation in the Fe-P bond-length (which is believed to affect the magnitude of the Fe moment \cite{che}) might
stabilize an incipient SDW in this material. However, there is no evidence for a magnetic or
density wave transition in single-crystals of LaFePO, and $T_c$ is uncorrelated with the
magnitude of the resistivity upturn, at least for the range of values that we have observed.
Hence, for such a scenario to be applicable, either the superconductivity must be insensitive
to the presence/absence of magnetic fluctuations, or there must be some degree of inhomogeneity
in the crystals, such that parts of the sample that suffer such fluctuations are physically
separate from the majority superconducting phase. Unfortunately, although this possibility is
especially interesting due to the parallels to LaFeAsO, we cannot definitively establish whether it is appropriate from these measurements. Additional experiments are underway to determine the
effect of stabilizing the SDW on the normal and superconducting states in this compound.

We note in closing that whatever the cause of the resistivity anomaly, it is reminiscent of an
upturn seen by McQueen {\it et al.} for polycrystalline samples which had no superconducting
transition.\cite{cava} The residual resistivity ratio of those polycrystalline samples was
nearly two orders of magnitude lower than that of the single-crystal samples described in this
report, and the resistivity upturn was significant. This raises the possibility that
the lack of superconductivity in those samples might not be due to differences in the oxygen
content, but rather to a larger concentration of defects, potentially affecting a competing
density wave ground state.

\section{CONCLUSIONS}

We have studied transport and thermal properties of single-crystals of \lafepo.
These samples exhibit superconductivity, and the evidence suggests that stoichiometric LaFePO is indeed a bulk superconductor, in contrast to other reports. Furthermore we find that the superconducting gap
is unrelated to non-magnetic disorder for values of $\xi/l\sim 0.2$(assuming that disorder is proportional to the interplane resistivity). This result is consistent with an $s$-wave pairing symmetry, but is not definitive. Finally, we have investigated an anomalous low-temperature resistivity upturn that
appears in some of our samples through detailed temperature and field studies. The feature
is related to disorder, though the present measurements are insufficient to definitively establish the origin of this effect.

\section{ACKNOWLEDGEMENTS}
We would like to thank Ross McDonald, Amalia Coldea, J. D. Fletcher, Scott Riggs, Stephen Kivelson, Doug Scalapino, Zach Fisk, Ben Powell, D. H. Lu, Ming Yi, Z. X. Shen and Theodore Geballe for useful
comments on this work prior to publication. This work is supported by the Department of Energy,
Office of Basic Energy Sciences under contract DE-AC02-76SF00515.


\begin{thebibliography}{1}
\bibitem{upturnnote} The temperature of 10K was chosen as the lower bound of our Fermi liquid
    fit because the resistivity lower than this and near $T_c$ is, for some samples, complicated
    by the presence of an up-turn.
\bibitem{dhva_comment} Due to electron counting arguments in combination with the Luttinger
    theorem the electron and hole FS pockets in the stoichiometric compound should enclose
    exactly equal volumes. Analysis of recent dHvA data taken for samples grown by the same
    technique described above indicate that the electron and hole pockets are almost perfectly
    balanced, in agreement with the anticipated result. Deviations from perfect charge balance
    in this analysis are slight, for instance corresponding to an oxygen deficiency of no more
    than 1.5$\%$. Orbits associated with the small 3D
    electron pocket centered at the Z point of the Brillouin zone have not been convincingly
    observed, and this may account for some of the deficiency, though unlikely to account for all.\cite{amalia}
\end{thebibliography}
\end{document}